\documentclass[preprintnumbers,superscriptaddress,amsmath,amssymb,longbibliography,preprint,nofootinbib]{revtex4-1}
\usepackage{graphicx}
\usepackage{amsfonts}
\usepackage[colorlinks,linkcolor=red,anchorcolor=blue,citecolor=green]{hyperref}
\usepackage{subfigure}
\usepackage{float}
\usepackage{enumerate}
\newcommand\diff{\mathrm{d}}

\makeatletter
\def\@bibdataout@aps{%
\immediate\write\@bibdataout{%
@CONTROL{%
apsrev41Control%
\longbibliography@sw{%
    ,author="08",editor="1",pages="1",title="0",year="1"%
    }{%
    ,author="08",editor="1",pages="1",title="",year="1"%
    }%
  }%
}%
\if@filesw \immediate \write \@auxout {\string \citation {apsrev41Control}}\fi 
}
\makeatother

\begin{document}
\title{
    Quasinormal modes of the spherical bumblebee black holes with a global monopole
}
\author{Rui-Hui Lin}
\email[]{linrh@shnu.edu.cn}
\author{Rui Jiang}
\author{Xiang-Hua Zhai}
\email[(corresponding author)]{zhaixh@shnu.edu.cn}
\affiliation{Division of Mathematics and Theoretical Physics, Shanghai Normal University, 100 Guilin Road, Shanghai 200234, China}

\begin{abstract}
    The bumblebee model is an extension of the Einstein-Maxwell theory
    that allows for the spontaneous breaking of the Lorentz symmetry of the spacetime.
    In this paper, we study the quasinormal modes of the spherical black holes in this model
    that are characterized by a global monopole.
    We analyze the two cases with a vanishing cosmological constant or a negative one (the anti-de Sitter case).
    We find that the black holes are stable under the perturbation of a massless scalar field.
    However, both the Lorentz symmetry breaking and the global monopole have notable impacts on the evolution of the perturbation.
    The Lorentz symmetry breaking may prolong or shorten the decay of the perturbation according to the sign of the breaking parameter.
    The global monopole, on the other hand, has different effects
    depending on whether a nonzero cosmological constant presences:
    it reduces the damping of the perturbations for the case with a vanishing cosmological constant,
    but has little influence for the anti-de Sitter case.
\end{abstract}
\maketitle
\newpage
\section{Introduction}
\label{intro}
One of the major challenges in theoretical physics is to reconcile Einstein's general relativity (GR),
the most successful theory of gravitation so far,
with the standard model of particle physics (SM),
which unifies all other interactions.
A possible clue to this problem is spontaneous symmetry breaking,
which plays a key role in the elementary particle physics.
In the early universe, the temperature may have been high enough to trigger symmetry breaking.
Depending on the topology of the manifold $M$ of degenerate vacua,
this process may result in different types of topological defects\cite{Kibble:1976sj,Vilenkin:1984ib}.
In particular, when second homotopy group $\pi_2M$ of $M$ is nontrivial,
point defects, i.e., monopoles are formed
and can be the source of inflation and seeds of the cosmic structures\cite{Barriola:1989hx,Bennett:1990xy}.
A simplest model to describe a global monopole is to consider
a triplet of scalar fields with a global O(3) symmetry spontaneously broken to U(1).
The gravitational field of a global monopole
describes a spacetime with a solid angle deficit\cite{Barriola:1989hx,Shi:1991yto}.
If such a global monopole is captured by an ordinary black hole,
a new type of black hole with a global monopole charge will be formed.
This charge makes the black hole topologically different from the usual one,
and may have interesting physical implications
(see, e.g., Refs. \cite{Gibbons:1990um,Horowitz:1999jd,Li:2002ku,Yu:2002st,Watabe:2003sxp,Chen:2009vz,Rizwan:2020znd,Soroushfar:2020wch,Luo:2022gss}).

Another possible symmetry breaking that may emerge in the quest of quantization of GR is the breaking of Lorentz symmetry.
It is shown that this symmetry may be strongly violated at the Planck scale ($\sim 10^{19}$ GeV) in some approaches to quantum gravity (QG)\cite{Mattingly:2005re,Amelino-Camelia:2008aez},
suggesting that it may not be a fundamental symmetry of nature.
Moreover, Lorentz symmetry breaking (LSB) may also provide candidates of observable signals of the underlying QG framework at low energy scale
\cite{Kostelecky:1988zi,Kostelecky:1991ak,Kostelecky:2003fs}.
Therefore, the occurrence of LSB has been explored in various scenarios,
such as string theory (see, e.g., \cite{Kostelecky:1988zi}),
loop quantum gravity\cite{Ellis:1999uh,Gambini:1998it},
massive gravity\cite{Fernando:2014gda}, Einstein-aether theory\cite{Jacobson:2007veq},
and others\cite{Arkani-Hamed:2003pdi,Burgess:2002tb,Cline:2003xy,Frey:2003jq}.
The effective field theory to study violations of fundamental symmetries in SM and gravity is called the Standard-Model Extension (SME)\cite{Carroll:2001ws,Lane:2019mgn}.
The gravitational couplings of SME is explored in a general Riemann-Cartan spacetime in Ref. \cite{Kostelecky:2003fs},
where it is shown that gravitational coupling spontaneous LSB can occur without geometrical incompatibilities.

The bumblebee model is a simple example to incorporate spontaneous LSB in the gravitational sector.
It is a vector-tensor theory that generalizes the Einstein-Maxwell theory by introducing a vector field $B_\mu$ that acquires a non-zero vacuum expectation value (VEV)
\cite{Kostelecky:1988zi,Kostelecky:1991ak,Kostelecky:2003fs,Bluhm:2004ep}.
This vector field $B_\mu$ couples nonminimally to the spacetime
and thus induces a spontaneous breaking of the local Lorentz symmetry
by selecting a preferred spacetime direction in the local frames through its frozen VEV.
When the bumblebee field is set to be a constant background,
it can be viewed as the coefficients related to LSB in SME\cite{Kostelecky:2009zp}.
The bumblebee model has attracted recent interests since an exact Schwarzschild-like black hole solution was found\cite{Casana:2017jkc}.
Other static\cite{Gullu:2020qzu,Maluf:2020kgf,Ding:2021iwv}
and rotating\cite{Ding:2019mal,Ding:2020kfr,Jha:2020pvk} solutions in the bumblebee model have been obtained.
The various physical aspects such as the gravitational lensing\cite{Ovgun:2018ran,DCarvalho:2021zpf},
shadow\cite{Casana:2017jkc,Ding:2019mal,Wang:2021irh}, accretion\cite{Liu:2019mls},
perturbations\cite{Kanzi:2021cbg,Oliveira:2021abg,Wang:2021gtd,Maluf:2022knd,Kanzi:2022vhp,Liu:2022dcn,Gogoi:2022wyv},
and superradiance\cite{Jiang:2021whw,Khodadi:2021owg,Khodadi:2022dff}, have also been studied.

In particular, by considering the bumblebee model with a global monopole,
G\"ull\"u and \"Ovg\"un constructed a family of Schwarzschild-like black hole solutions
and analyzed the influence of the LSB parameter on the black hole shadow and the weak deflection angle\cite{Gullu:2020qzu}.
Furthermore, in the study of the generalized uncertainty principle correction of the bumblebee black holes,
Gogoi and Goswami proposed that a similar family of black hole solutions
with both a global monopole and a non-vanishing cosmological constant can also be constructed\cite{Gogoi:2022wyv}.
These two families of black holes, as strong field solutions,
provide optimal environments for investigating spontaneous LSB and global monopoles within the gravitational context.
When black holes are perturbed, the resulting oscillations and their evolution may reveal unique characteristics of the black hole,
or equivalently, the spacetime and gravity.
These oscillating modes are referred to as quasi-normal modes (QNMs)
and have become a widely utilized tool for analyzing the stability of black hole spacetime
since their use in demonstrating the stability of the Schwarzschild black hole\cite{Regge:1957td,Vishveshwara:1970cc}.
In the current paper, we focus on the QNMs of aforementioned two families of black hole solutions,
namely the spherical bumblebee black holes with a global monopole in the presence or absence of a cosmological constant.
Our aim is to study how the spontaneous LSB and the global monopole affect the stability of the black hole spacetime.

This paper is structured as follows.
In Section \ref{review}, we provide a brief overview of spherical black holes in the bumblebee model with a global monopole
and establish the configuration of perturbation by considering an impinging massless scalar field.
In Section \ref{QNMs}, we examine the QNMs of these spherical bumblebee black holes with a global monopole for vanishing and non-vanishing cosmological constant.
Our study is concluded in Section \ref{conclusion}.
Throughout the paper, we follow the metric convention $\left(+,-,-,-\right)$  and use the units $G=\hbar=c=1$.

\section{Spherical bumblebee black holes with a global monopole and their scalar perturbation}
\label{review}
\subsection{Spherical bumblebee black holes with a global monopole}
The Lagrangian of the bumblebee model is generally written as\cite{Casana:2017jkc,Maluf:2020kgf}
\begin{equation}\label{lag}
    \mathcal{L} _{B}=\sqrt{-g} \left[
    \frac{1}{16\pi } \left ( R-2\Lambda+\xi B^{\mu }B^{\nu } R_{\mu \nu }   \right ) -\frac{1}{4} B^{\mu \nu }B_{\mu \nu }-V\left ( B_{\mu }B^{\mu}\pm b^2 \right ) \right]+\mathcal{L} _{M},
\end{equation}
where $B_{\mu\nu}$ denotes the strength of the bumblebee field $B_\mu$ and is defined as
\begin{equation}
    B_{\mu \nu }=\partial _{\mu } B_{\nu}-\partial _{\nu} B_{\mu}.
\end{equation}
The term with the parameter $\xi$ represents the quadratic coupling between the bumblebee field and the Ricci tensor $R_{\mu\nu}$.
$\Lambda$ is the cosmological constant and $\mathcal L_M$ is the Lagrangian of matter.
The term $V(B_{\mu }B^{\mu}\pm b^2)$ refers to the potential of the bumblebee field that takes a minimum at $B_{\mu }B^{\mu}\pm b^2=0$ with $b^2>0$.
The $\pm$ signs here associate to a spacelike or timelike bumblebee field, respectively.
For a stable vacuum spacetime, one would expect that the potential $V$ will be minimized
and the bumblebee field $B_\mu$ will be frozen at the VEV $\langle B_{\mu}\rangle =b_{\mu }$ with $b_\mu b^\mu=\mp b^2$.
Due to the coupling featured by the parameter $\xi$, such a nonzero VEV of $B_\mu$ then will break Lorentz symmetry
by selecting a preferred direction of spacetime.
Variation of the Lagrangian \eqref{lag} with respect to the metric $g^{\mu\nu}$ gives the equation of motion
\begin{equation}\label{eom}
    \begin{aligned}
        0= & R_{\mu \nu}+\Lambda g_{\mu\nu}-\kappa \left(T_{\mu \nu }^{(M)}-\frac{1}{2}g_{\mu \nu }T^{(M)} \right) +\xi b_{\mu } b^{\alpha }R_{\alpha \nu }                                              \\
           & +\xi b_{\nu } b^{\alpha }R_{\alpha \mu }-\frac{\xi }{2} g_{\mu \nu } b^{\alpha }b^{\beta}R_{\alpha\beta}-\frac{\xi }{2} \nabla _{\alpha }\nabla _{\mu} \left ( b^{\alpha }b_{\nu } \right ) \\
           & -\frac{\xi }{2} \nabla _{\alpha }\nabla _{\nu} \left ( b^{\alpha }b_{\mu } \right )+\frac{\xi }{2} \nabla^{2}\left ( b_{\mu } b_{\nu }  \right ) +\kappa (g_{\mu \nu}b^2-2b_\mu b_\nu )V',
    \end{aligned}
\end{equation}
where $\kappa=8\pi$ and $V'$ denotes $\diff V(x)/\diff x$.
For a global monopole field, the energy-momentum tensor $T_{\mu \nu }^{(M)}$ has the form\cite{Barriola:1989hx}
\begin{equation}\label{T}
    T_{\mu}^{(M)\nu }=\text{diag}\left(\frac{\eta ^{2}}{r^{2}},\frac{\eta ^{2}}{r^{2}},0,0\right),
\end{equation}
where $\eta$ is a constant related to the global monopole charge.
For the usual spherical ansatz of metric
\begin{equation}
    \label{sphansatz}
    \diff s^2=f(r)\diff t^2-\frac{\diff r^2}{h(r)}-r^2\diff\theta^2-r^2\sin^2\theta\diff\phi^2
\end{equation}
where $f(r)$ and $h(r)$ are the metric functions of the radial coordinate $r$,
it is assumed that the bumblebee field has a purely radial form $b_{\mu}=(0,b_{r}(r),0,0)$,
and hence, $B_\mu B^\mu=-b_r^2$.

In the scenario with a vanishing $\Lambda$,
a spherical solution can be constructed when the potential $V$ satisfies $V'=0$
at its minimum where $B^{\mu} B_{\mu }+b^2=0$\cite{Casana:2017jkc,Gullu:2020qzu}.
This solution is shown to be Schwarzschild-like with a global monopole and can be given by\cite{Gullu:2020qzu}
\begin{equation}\label{schlike}
    \begin{split}
        f(r)=   & 1-\kappa \eta ^{2}-\frac{2M}{r}, \\
        h(r)=   & \frac{L+1}{f(r)},\\
        b_r(r)= & \frac{|b|}{\sqrt{h(r)}},
    \end{split}
\end{equation}
where $L\equiv\xi b^2$ is the LSB parameter,
and $M$ is an integration constant corresponding to the Arnowitt-Deser-Misner mass.
The event horizon can be easily found to be at $r_h=2M/(1-\kappa\eta^2)$,
independent of the bumblebee field.

For the case where the cosmological constant $\Lambda$ is nonzero,
it is suggested that solutions can be constructed by setting\cite{Maluf:2020kgf}
\begin{equation}\label{VB}
    V\left ( B^{\mu} B_{\mu }+b^2 \right ) =\frac{\lambda }{2} \left ( B^{\mu} B_{\mu }+b^2 \right ),
\end{equation}
and viewing this term as a Lagrange multiplier term that ensures $B_\mu B^\mu+b^2=0$ in the variation of the Lagrangian \eqref{lag}.
Following this scheme,
we check that the Schwarzschild-(a)dS-like bumblebee black hole solution with a global monopole
proposed in Ref. \cite{Gogoi:2022wyv} can be constructed, which is given by
\begin{equation}\label{schadslike}
    \begin{split}
        f(r)=   & 1-\kappa \eta ^{2}-\frac{2M}{r}-\frac\Lambda3r^2, \\
        h(r)=   & \frac{L+1}{f(r)},\\
        b_r(r)= & \frac{|b|}{\sqrt{h(r)}},\\
        \lambda=&\frac{\xi\Lambda}{\kappa\left(L+1\right)}.
    \end{split}
\end{equation}

\subsection{Scalar perturbation}
We study the perturbation of the spherical bumblebee black holes \eqref{schlike} and \eqref{schadslike}
by considering an impinging massless scalar field $\Phi$.
We restrict our attention to the cases with $\Lambda\le0$,
as the cases with $\Lambda>0$ have a cosmological horizon that limits the causal structure of the spacetime.

We assume that $\Phi$ is minimally coupled to gravity and satisfies the covariant Klein-Gordon equation,
\begin{equation}\label{K-G}
    \frac{1}{\sqrt{-g}}\partial _{\mu }\left(g^{\mu\nu}\sqrt{-g}\partial_{\nu}\Phi \right)=0.
\end{equation}
Using the separation of $\Phi$ in the spherical coordinates $(t,r,\theta,\phi)$,
\begin{equation}\label{separation1}
    \Phi=\frac{\psi_{l}(r)}{r}Y_{lm}(\theta ,\phi ) e^{-i\omega t},
\end{equation}
where $Y_{lm}(\theta ,\phi )$ is the spherical harmonic function, we can obtain the radial equation of Eq. \eqref{K-G} as
\begin{equation}\label{radial}
    \frac{1}{1+L} f\frac\diff{\diff r}\left(f\frac{\diff\psi_{l}}{\diff r}\right)+\left [  \omega ^{2}-U(r) \right ]\psi_{l}=0.
\end{equation}
The effective potential $U(r)$ is defined as
\begin{equation}\label{potential}
    U(r)=f(r)\left [ \frac{l(l+1)}{r^2} +\frac{f'(r)}{(1+L)r}  \right ].
\end{equation}
Introducing the tortoise coordinate $x$ defined by
\begin{equation}\label{toordinate}
    \dfrac{\diff x}{\diff r}=\frac{\sqrt{1+L} }{f(r)},
\end{equation}
one can rewrite the radial equation \eqref{radial} as
\begin{equation}\label{radial2}
    \frac{\diff^2 \psi_{l}}{\diff x^2} +\left [  \omega ^{2}-U(x) \right ]\psi_{l}=0.
\end{equation}

To keep track of the temporal evolution of $\Phi$ that is encoded in $\omega$ in the radial equation \eqref{radial},
one can, alternatively, adopt a different separation of $\Phi$ as
\begin{equation}
    \Phi=\frac{\Psi(t,r)}{r}Y_{lm}(\theta ,\phi ),
\end{equation}
where its dependence on $r$ and $t$ are kept in a single function $\Psi(t,r)$.
Then, the equation for $\Psi$ from Eq. \eqref{K-G} is
\begin{equation}\label{radial3}
    \left [ \frac{\partial ^2}{\partial x^2} -\frac{\partial ^2}{\partial t^2} -U(x) \right ] \Psi(x,t)=0.
\end{equation}
With the light cone coordinates
\begin{equation}\label{light cone coordinates}
    du=dt-dx,\qquad dv=dt+dx,
\end{equation}
one can rewrite Eq.\eqref{radial3} as
\begin{equation}\label{radial4}
    \left [ 4\frac{\partial^2}{\partial u\partial v} +U(u,v) \right ] \Psi(u,v)=0.
\end{equation}

\begin{figure}
    \centering
    \includegraphics[width=0.48\linewidth]{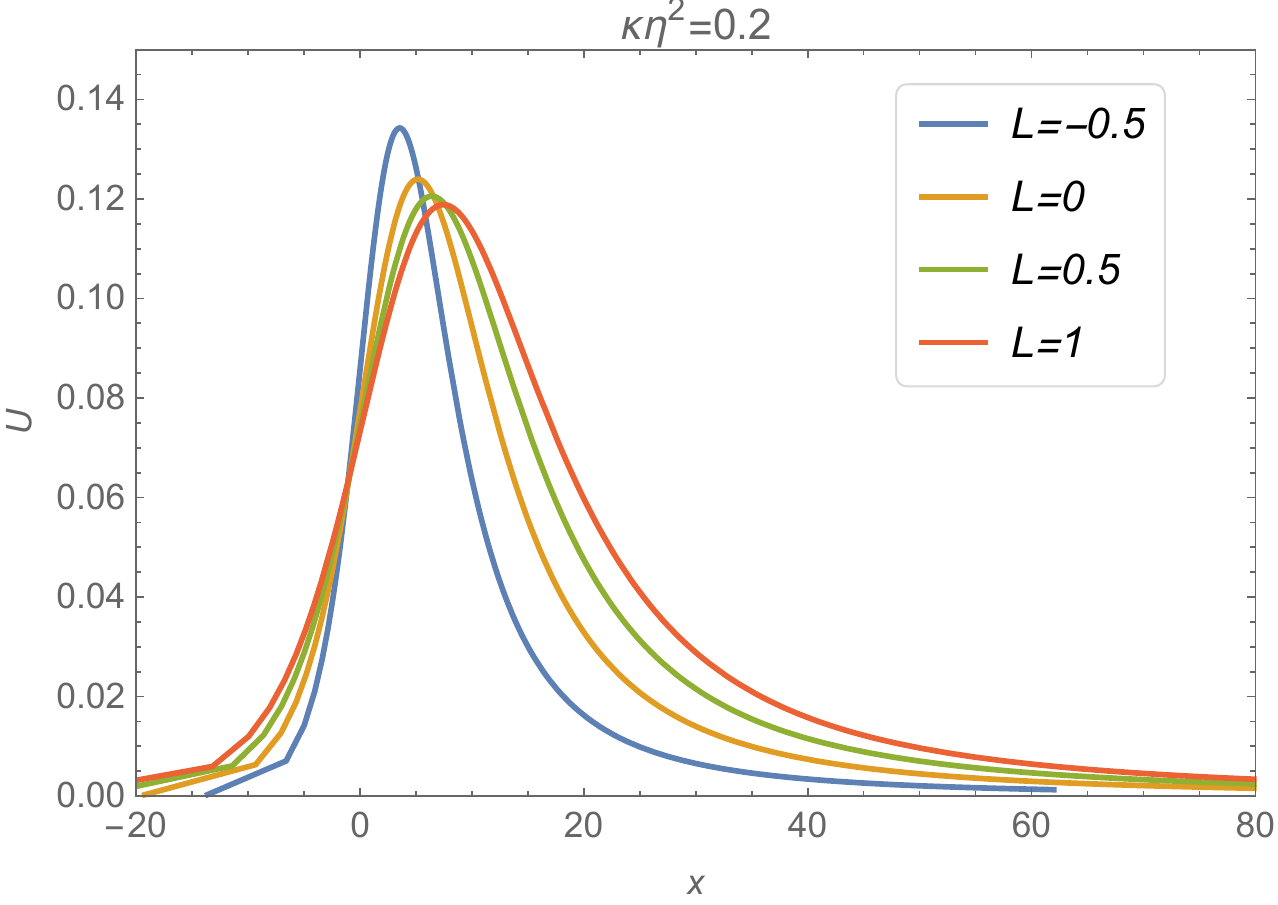}
    \includegraphics[width=0.48\linewidth]{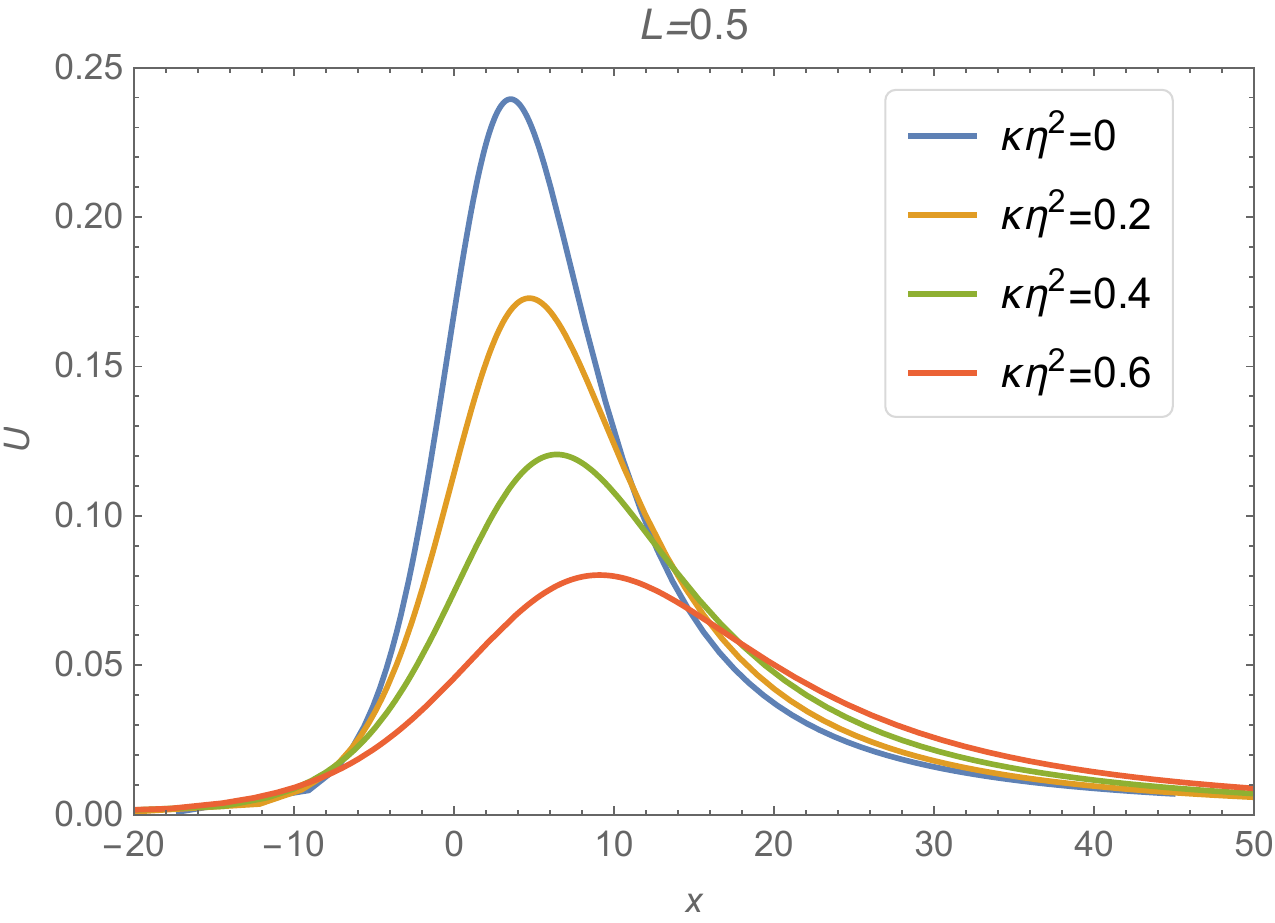}
    \caption{The effective potential for scalar perturbations with different $L$
        and $\kappa\eta^2$ when $l=2$, where we have set $M=1$ and $\Lambda=0$.}
    \label{schlike_effU}
\end{figure}

\begin{figure}
    \centering
    \includegraphics[width=0.48\linewidth]{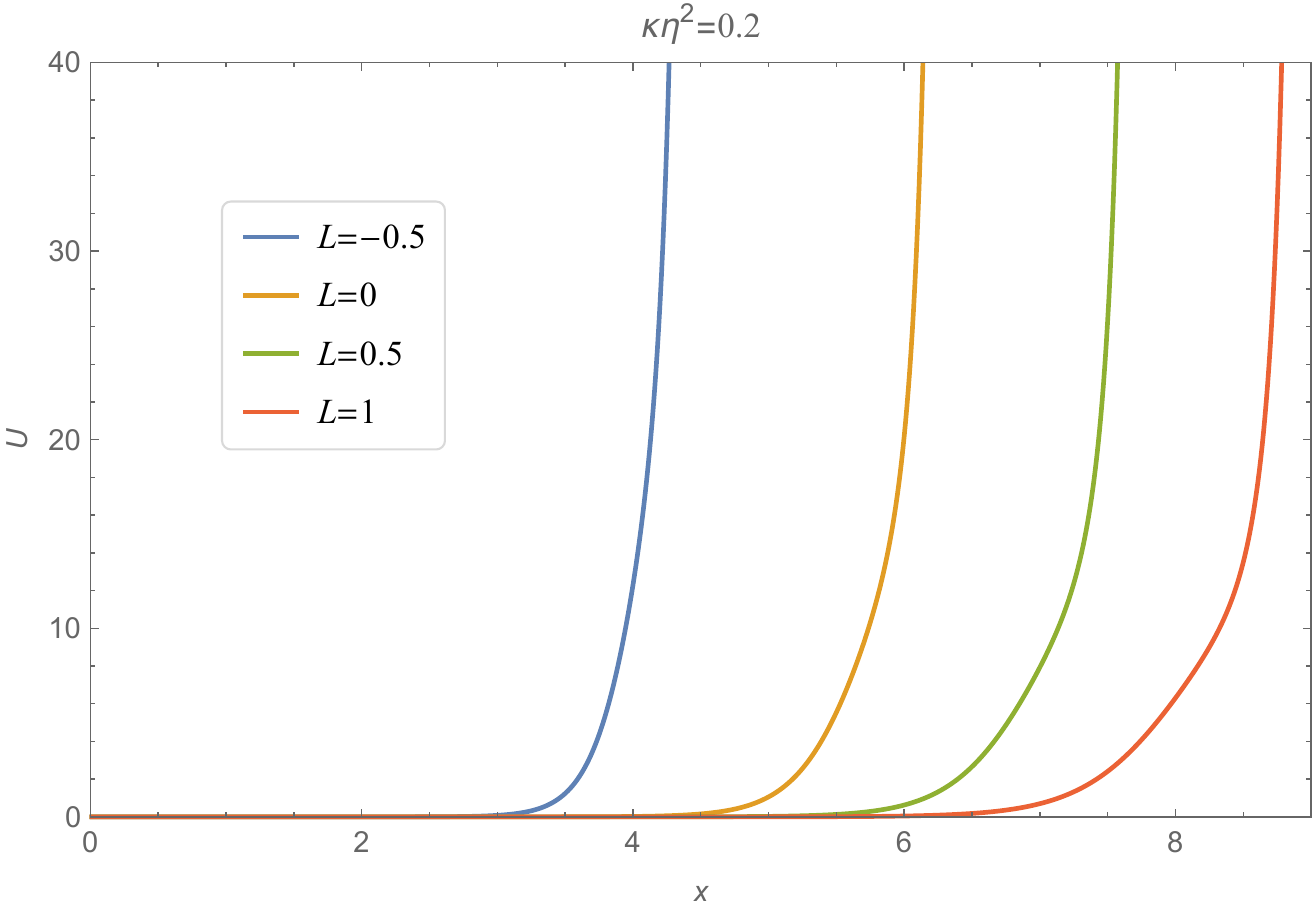}
    \includegraphics[width=0.48\linewidth]{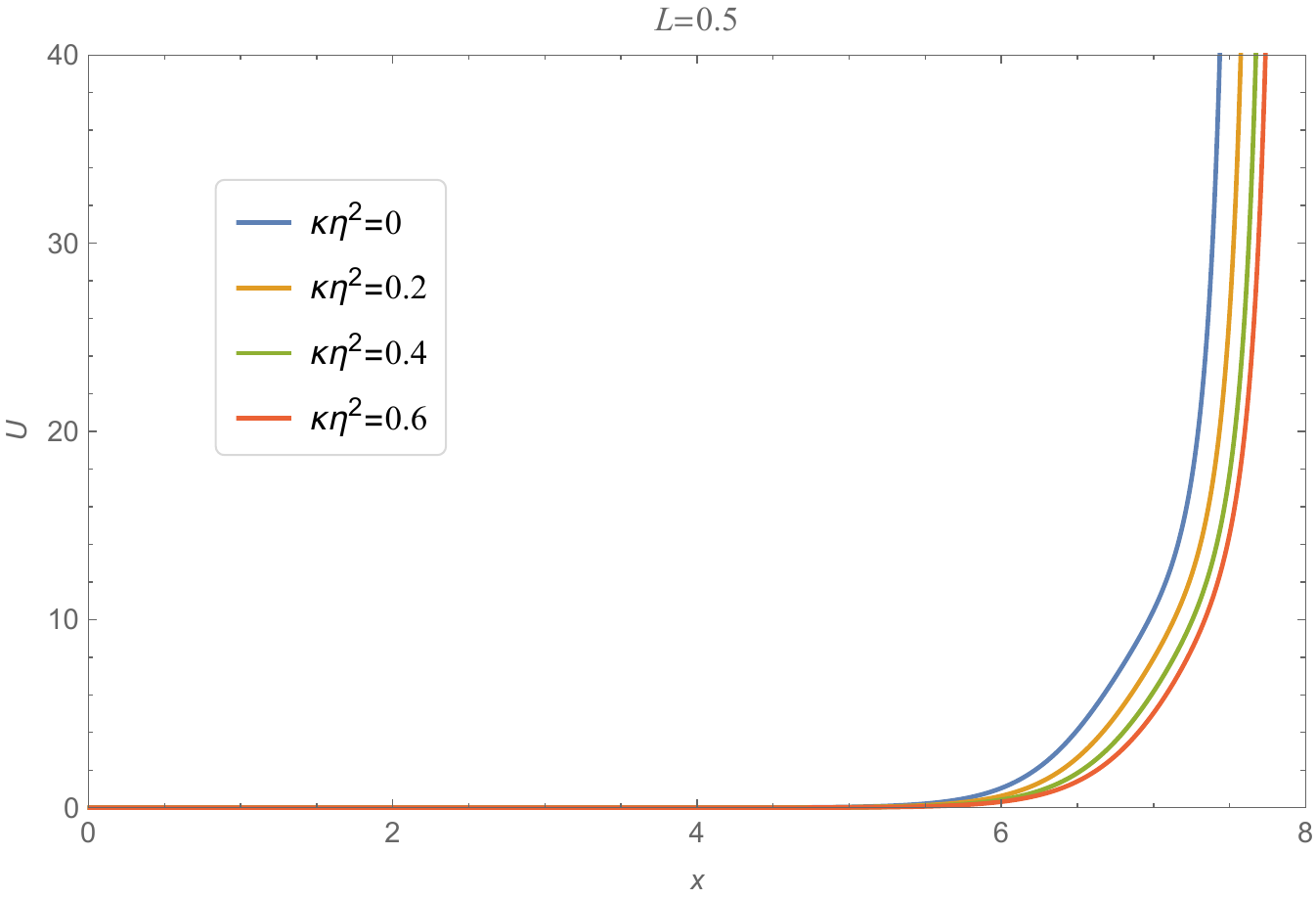}
    \caption{The effective potential for scalar perturbations with different $L$
        and $\kappa\eta^2$ when $l=2$, where we have set $M=1$ and $\Lambda=-3$.}
    \label{schads_effU}
\end{figure}

Generally, based on different separations, either Eq. \eqref{radial2} or \eqref{radial4} can be solved numerically
once the boundary conditions are set,
which we will do for the bumblebee black holes in the next section.
Here we concentrate on the effective potential $U$ defined in Eq. \eqref{potential}.
The behaviors of $U$
for vanishing and negative $\Lambda$ are shown in Figs. \ref{schlike_effU} and \ref{schads_effU}, respectively.
The tortoise coordinate $x$ ranges from $-\infty$ at the event horizon
to $+\infty$ at spatial infinity.
For $\Lambda=0$,
the effective potential tends to zero both near the horizon
and at large distances from the black hole.
For $\Lambda<0$, however,
the effective potential diverges at spatial infinity.
The effects of the LSB parameter $L$ and
the global monopole parameter $\kappa\eta^2$
on the effective potential are evident.

\section{QNMs of the bumblebee black holes with a global monopole}
\label{QNMs}
\subsection{The black hole with a vanishing cosmological constant}
As seen in the previous section,
the effective potential $U$ for the scalar field approaches zero at $x\to\pm\infty$
in this case.
Therefore, together with the physical consideration that
the field should be purely outgoing towards spatial infinity
and purely ingoing at the horizon,
the asymptotic forms of the field can be written as
\begin{equation}\label{bc1}
    \psi_{l}\sim \left\{
    \begin{aligned}
         & e^{-i\omega x }\quad & ,\quad x\to -\infty , \\
         & e^{i\omega x } \quad & ,\quad x\to +\infty.
    \end{aligned}
    \right.
\end{equation}
The radial equation \eqref{radial2} and the boundary condition \eqref{bc1}
determine a discrete spectrum of frequencies $\{\omega_n\}$
for a fixed pair of symmetry breaking parameters $L$ and $\kappa\eta^2$.
We use the continued-fraction method\cite{Leaver:1985ax} to solve this problem numerically.
The frequency $\omega$ is generally complex,
with a real part $\omega_R$ corresponding to the oscillation frequency of the field
and an imaginary part $\omega_I$ reflecting the evolution of the wave amplitude.
A positive $\omega_I$ implies a growing scalar field with a growth rate of $\omega_I$.
A negative $\omega_I$ implies a decaying scalar field with a damping rate of $|\omega_I|$.

\begin{figure}
    \centering
    \includegraphics[width=0.48\linewidth]{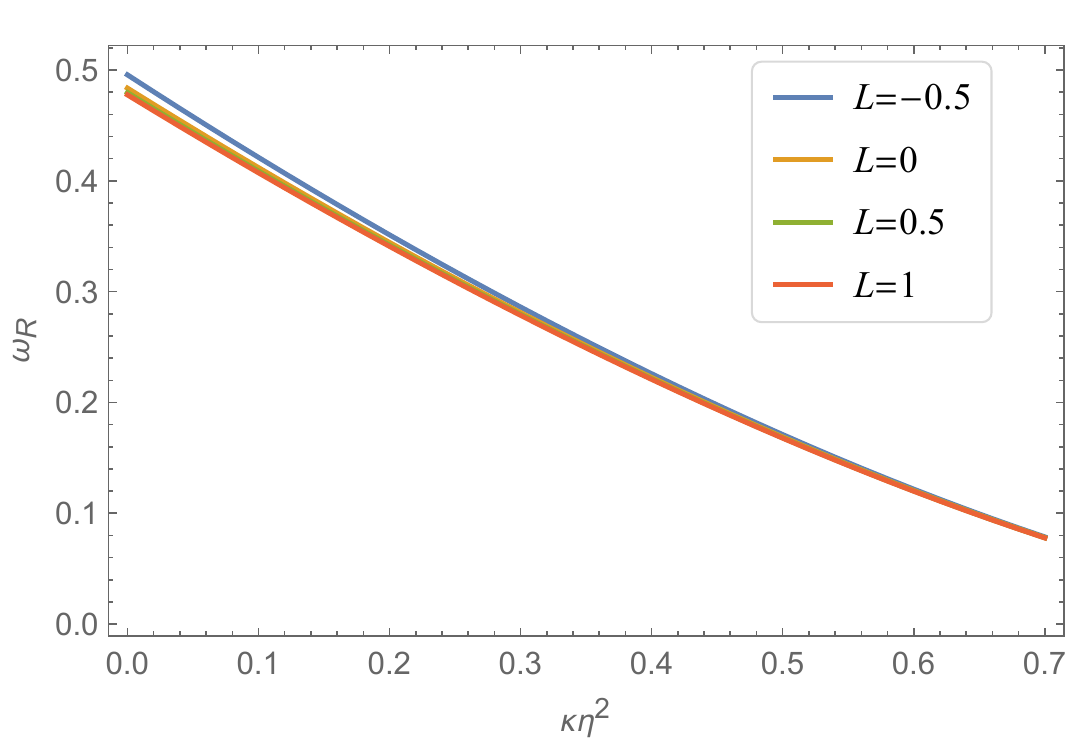}
    \includegraphics[width=0.48\linewidth]{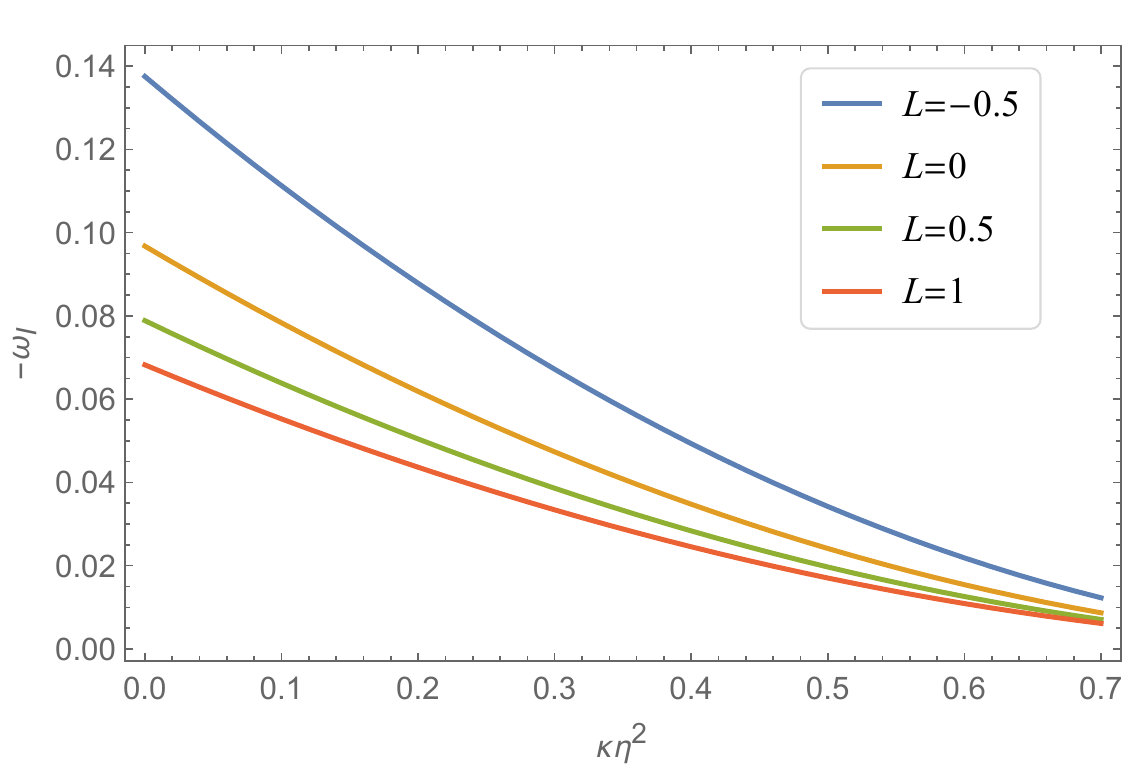}
    \caption{The real part (left) and imaginary part (right) of QNM frequencies
        of the spherical bumblebee black holes with a global monopole and a vanishing $\Lambda$,
        where we have set $M=1, l=2$ and $n=0$.}
    \label{schlikefreq}
\end{figure}
The QNM frequencies as functions of $\kappa\eta^2$ for different $L$ are shown in Fig. \ref{schlikefreq}.
The left panel shows that the real part of the frequencies, $\omega_R$,
decreases with increasing $L$ or $\kappa\eta^2$.
The right panel shows that the imaginary part of the frequencies, $\omega_I$,
is negative for all values of $L$ and $\kappa\eta^2$,
indicating that the black hole is stable under the perturbation of a massless scalar field.
However, we also observe that $|\omega_I|$ becomes smaller as $L$ or $\kappa\eta^2$ increases.
This implies that for the black hole with larger $L$ or $\kappa\eta^2$,
the perturbation takes longer time to decay away.
The effect of $L$ on the stability in this case agrees with
the results of other bumblebee black holes reported in Refs. \cite{Kanzi:2021cbg,Oliveira:2021abg}.
We note that $\kappa\eta^2\ge0$, so the perturbation to the black hole
without a global monopole ($\kappa\eta^2=0$) fades out more quickly
than the case with one ($\kappa\eta^2>0$).

\begin{figure}
    \centering
    \includegraphics[width=0.48\linewidth]{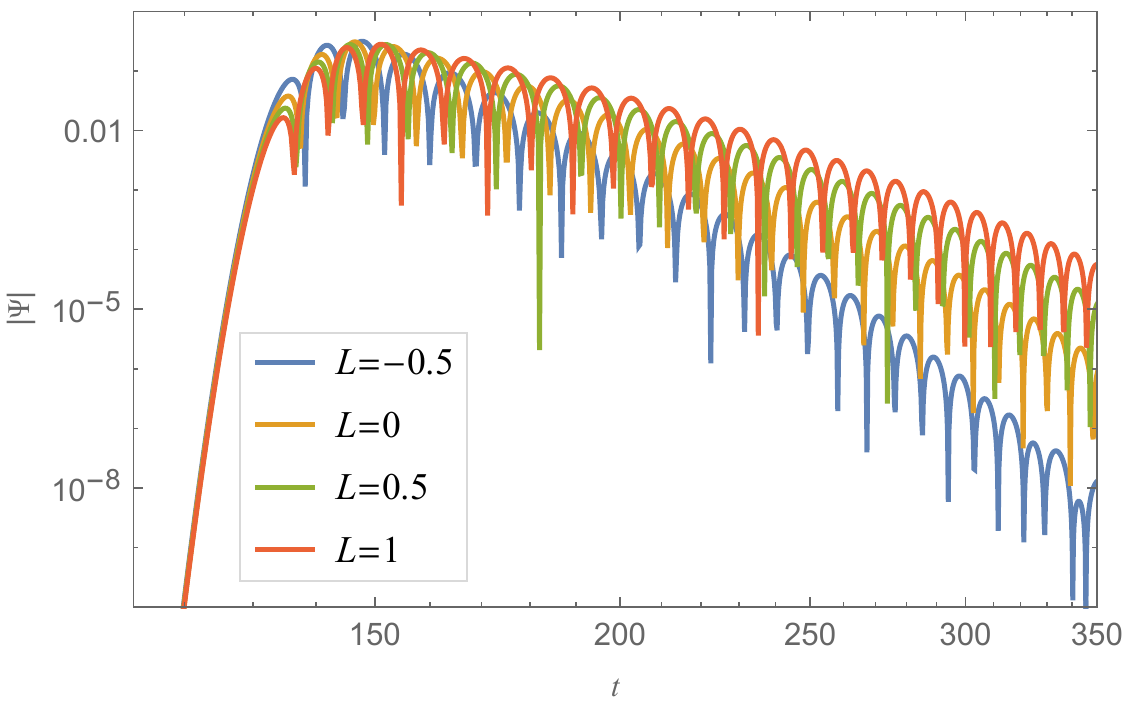}
    \includegraphics[width=0.48\linewidth]{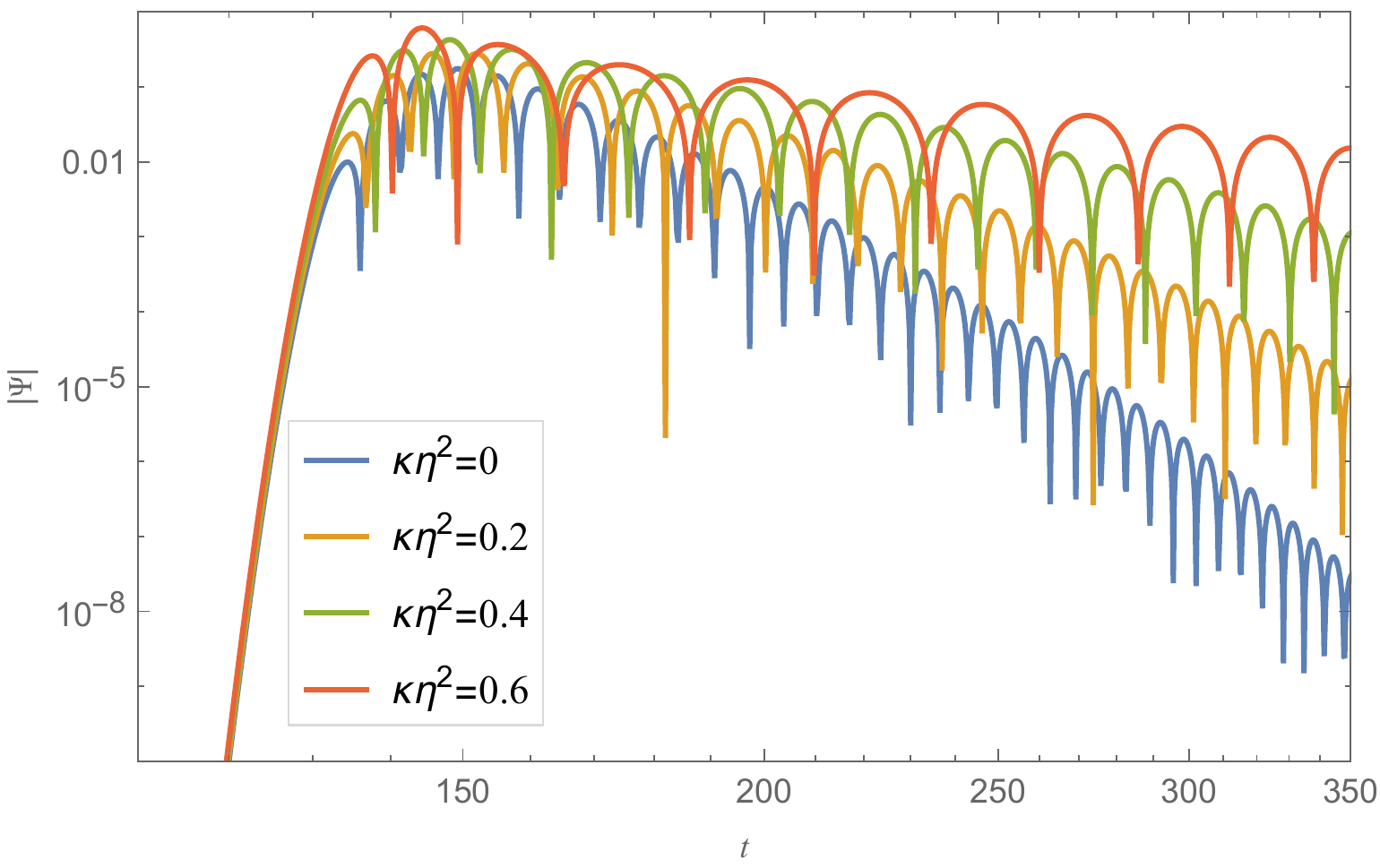}
    \caption{Temporal evolution of the scalar perturbations,
        with $\kappa \eta ^{2}=0.2,l=2$ in the left panel
        and $L=0.5,l=2$ in the right panel.}
    \label{temp-sch}
\end{figure}

We also investigate the temporal evolution of the perturbation by solving Eq. \eqref{radial4}.
We set the initial conditions as
\begin{equation}\label{bc2}
    \Psi(u,0)=0,\qquad  \Psi (0,v)=1.
\end{equation}
The numerical results are displayed in Fig. \ref{temp-sch}.
They confirm that the black hole is stable under the massless scalar field perturbation.
The left panel shows the decay of the perturbation for different $L$ with $\kappa\eta^2=0.2$ and $l = 2$,
in which it is obvious that the scalar field decays slower for larger $L$.
The right panel shows the decay of the perturbation for different $\kappa\eta^2$ with $L = 0.5$ and $l = 2$,
where one can see that the scalar field decays slower for larger $\kappa\eta^2$.
This agrees with the results in the frequency domain.
Another consistent observation from Figs. \ref{schlikefreq} and \ref{temp-sch} is that
the oscillation frequencies are more sensitive to the monopole charges $\eta^2$ than to the LSB parameters $L$.

\begin{figure}
    \centering
    \includegraphics[width=0.48\linewidth]{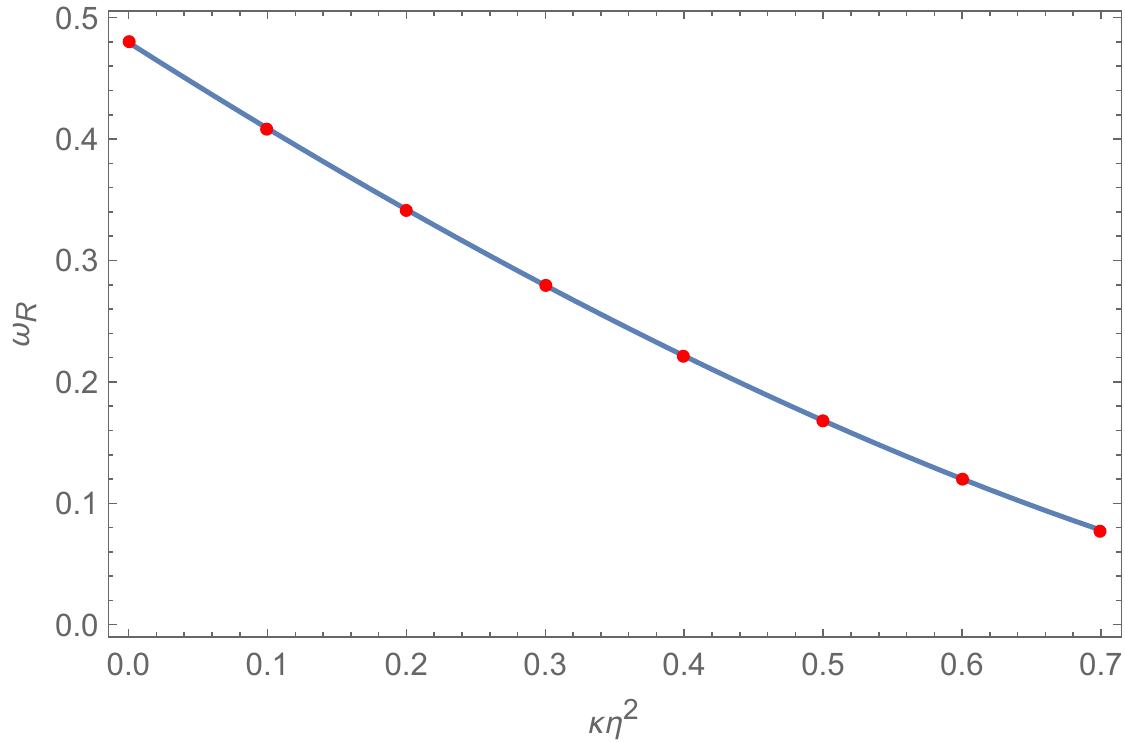}
    \includegraphics[width=0.48\linewidth]{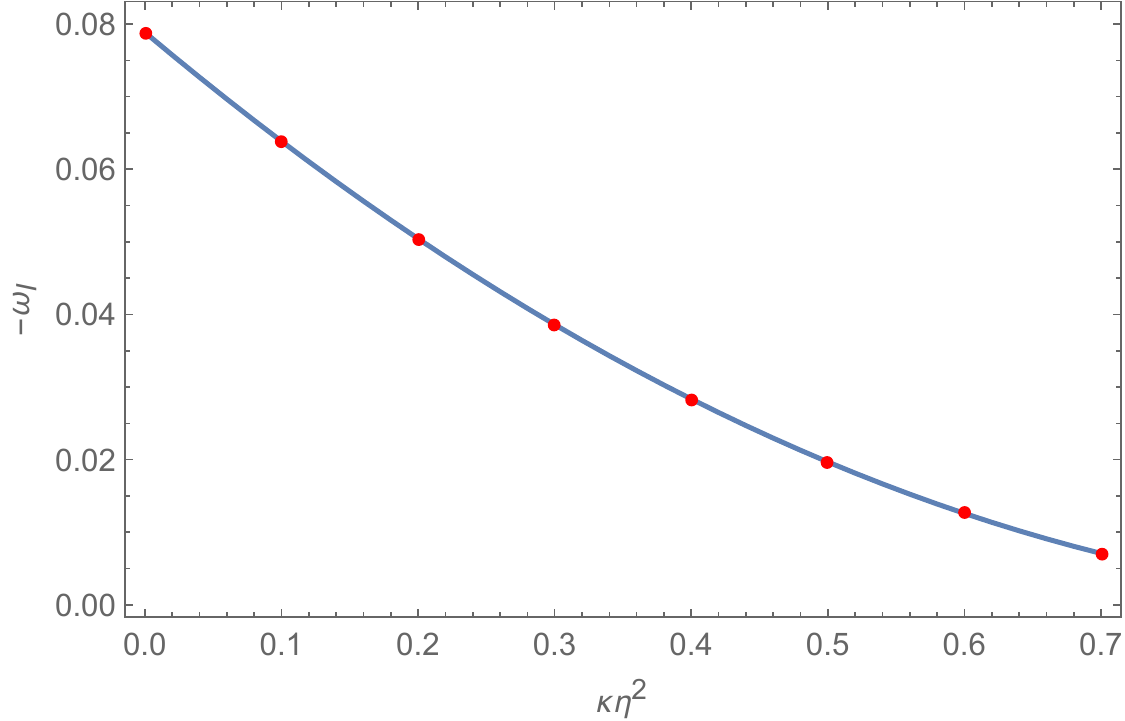}
    \caption{Comparison of the QNM frequencies obtained by the continued-fraction method (blue line)
        and those extracted from time domain analysis (red dots), with $M=1, L=0.5,l=2$.}
    \label{prony-sch}
\end{figure}

We use the Prony method\cite{Berti:2007dg} to extract the QNM frequencies from the time domain profile
and compare them with the results from the continued-fraction method.
For instance, we find the frequency
$\omega=0.342031-0.0504228i$ for $\kappa \eta ^{2}=0.2,L=0.5,l=2$ by using the Prony method.
The continued-fraction method gives $\omega=0.342003-0.0504359i$ for the same case,
which has an error less than one thousandth.
Fig. \ref{prony-sch} shows the comparison of the QNMs obtained by the two methods for different $\kappa\eta^2$,
which demonstrates a good agreement between the two methods.

\subsection{The black hole with a negative cosmological constant}
\begin{figure}
    \centering
    \includegraphics[width=0.48\linewidth]{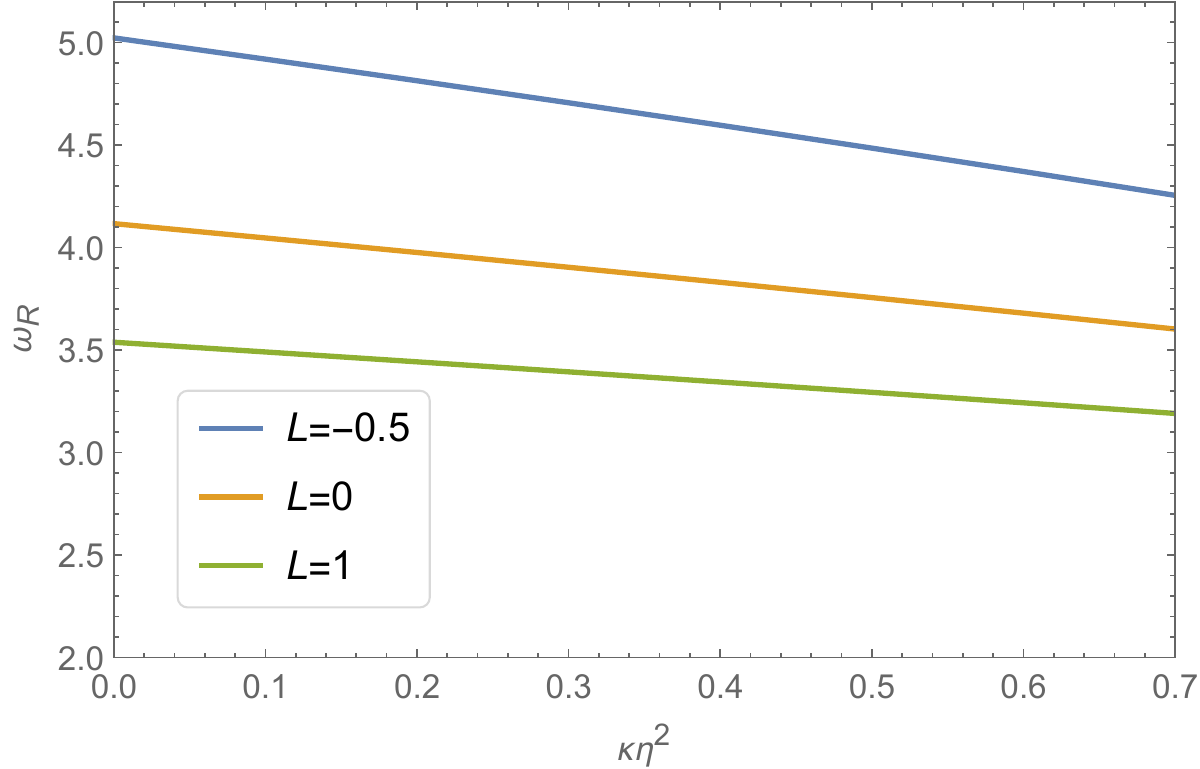}
    \includegraphics[width=0.48\linewidth]{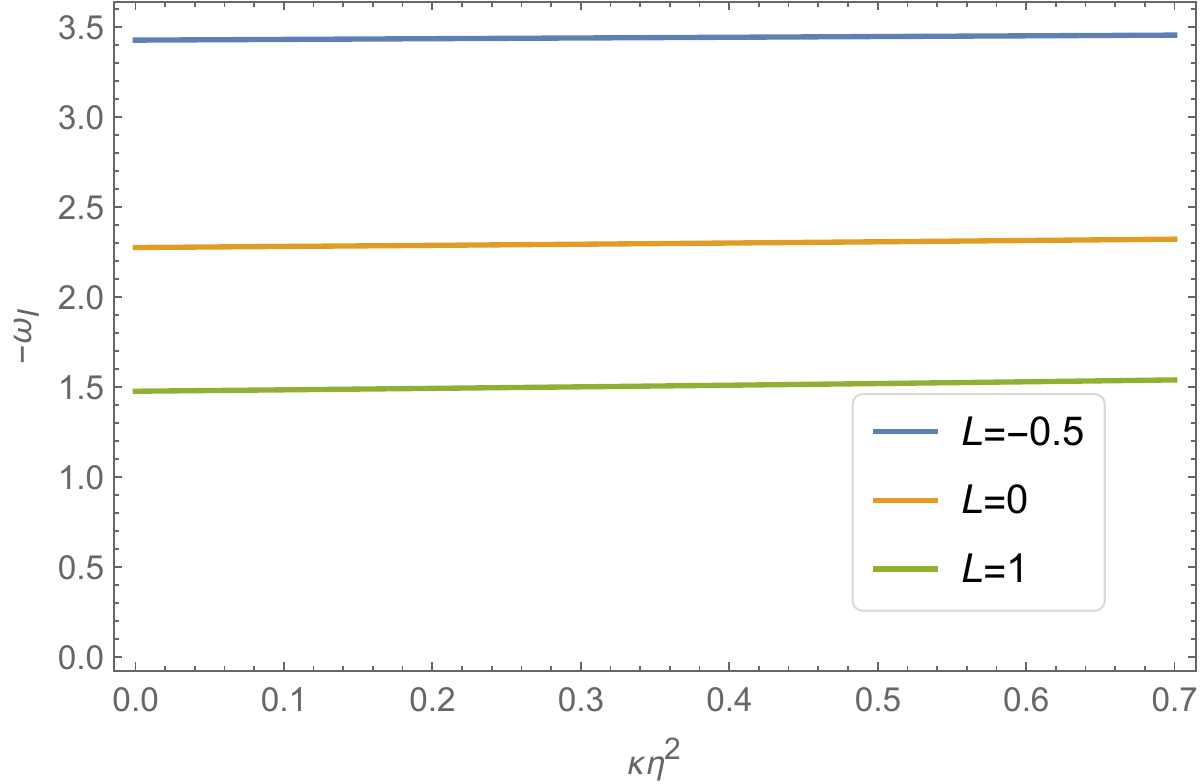}
    \caption{The real part (left) and imaginary part (right) of QNM frequencies
        of the spherical bumblebee black holes with a global monopole and non-zero $\Lambda$,
        where we have set $\Lambda=-3, l=2, n=0$
        and a suitable $M$ such that the horizon radius $r_+=1$.}
    \label{Fig.QNM-ads}
\end{figure}

For $\Lambda<0$,
the effective potential $U$ will diverge at spatial infinity,
as shown in Fig. \ref{schads_effU}.
This requires that $\Phi=0$ at spatial infinity.
Therefore, the asymptotic forms of $\Psi$ in this case should be
\begin{equation}\label{boundary condition AdS}
    \Psi\sim \left\{
    \begin{aligned}
         & e^{-i\omega x }\quad & ,\quad x\to -\infty , \\
         & 0\quad               & ,\quad x\to +\infty.
    \end{aligned}
    \right.
\end{equation}
We apply the Horowitz-Hubeny method\cite{Horowitz:1999jd} to solve Eq. \eqref{radial2}
for the spectrum of frequencies with these Dirichlet boundary conditions.
The resulting QNM frequencies as functions of $\kappa\eta^2$ for various $L$ are depicted in Fig. \ref{Fig.QNM-ads}.
The left panel reveals that $\omega_R$ decreases with increasing $L$ or $\kappa\eta^2$,
implying that the scalar field oscillates at a lower frequency for larger $L$ or $\kappa\eta^2$.
The right panel demonstrates that $\omega_I$ is negative for all values of $L$ and $\kappa\eta^2$,
indicating the stability of the bumblebee black hole
under massless scalar field perturbations.
Furthermore, we observe that $|\omega_I|$ decreases with increasing $L$,
suggesting that the perturbation to the black hole takes longer time to decay away for larger $L$.
On the other hand, $|\omega_I|$ is insensitive to $\kappa\eta^2$,
which differs significantly from the zero-$\Lambda$ scenario.

\begin{figure}
    \centering
    \includegraphics[width=0.5\linewidth]{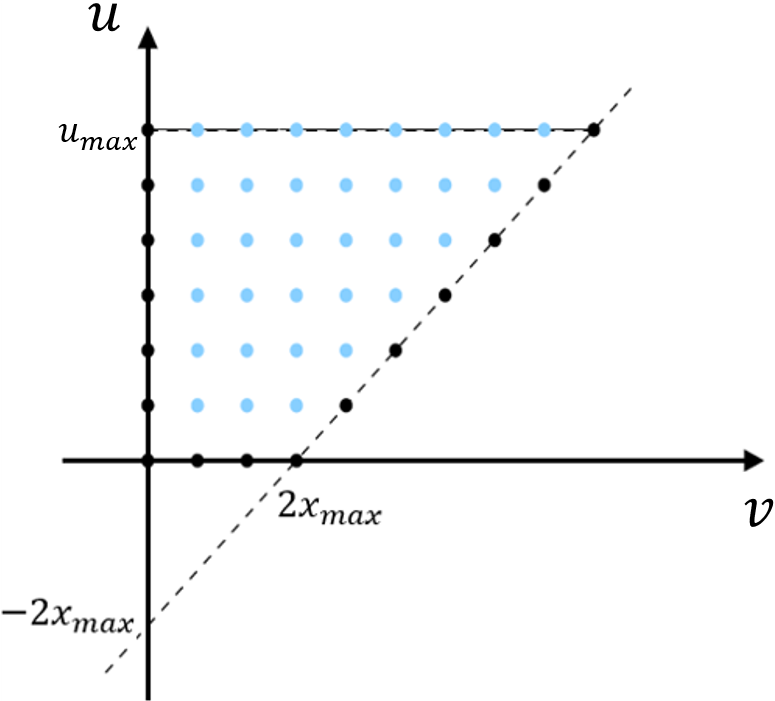}
    \caption{Schematic of the numerical grid and the calculation domain.
        The black dots denote the grid points where the field values are given by the initial condition and the boundary condition.
        The blue dots denote the grid points to be computed.\cite{Wang:2000dt}}
    \label{uvarea}
\end{figure}

To compute the temporal evolution of the QNMs in this setting,
we need to limit the calculation at a finite $x=x_\text{max}$,
since the potential diverges at spatial infinity.
This implies that the calculation domain on the $u$-$v$ plane
is bounded by the line $v-u=2x_\text{max}$\cite{Wang:2000dt}, as illustrated in Fig. \ref{uvarea}.
Hence, we can impose the boundary condition
\begin{equation}
    \Psi(v-u=2x_\text{max})=0.
\end{equation}

\begin{figure}
    \centering
    \includegraphics[width=0.48\linewidth]{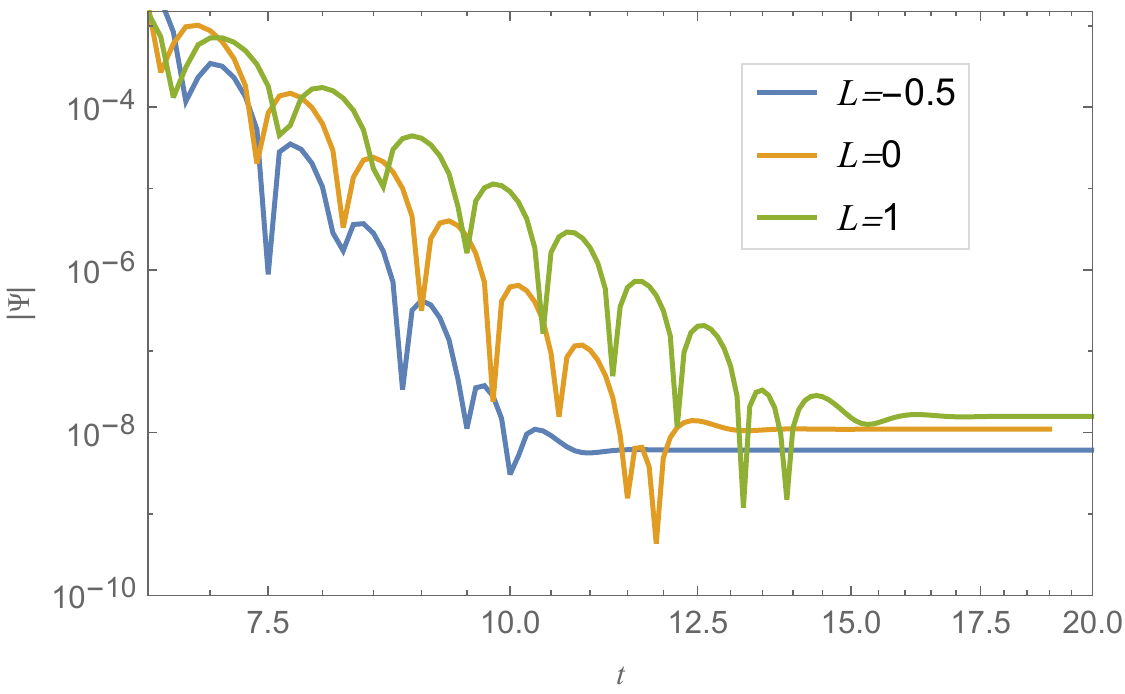}
    \includegraphics[width=0.48\linewidth]{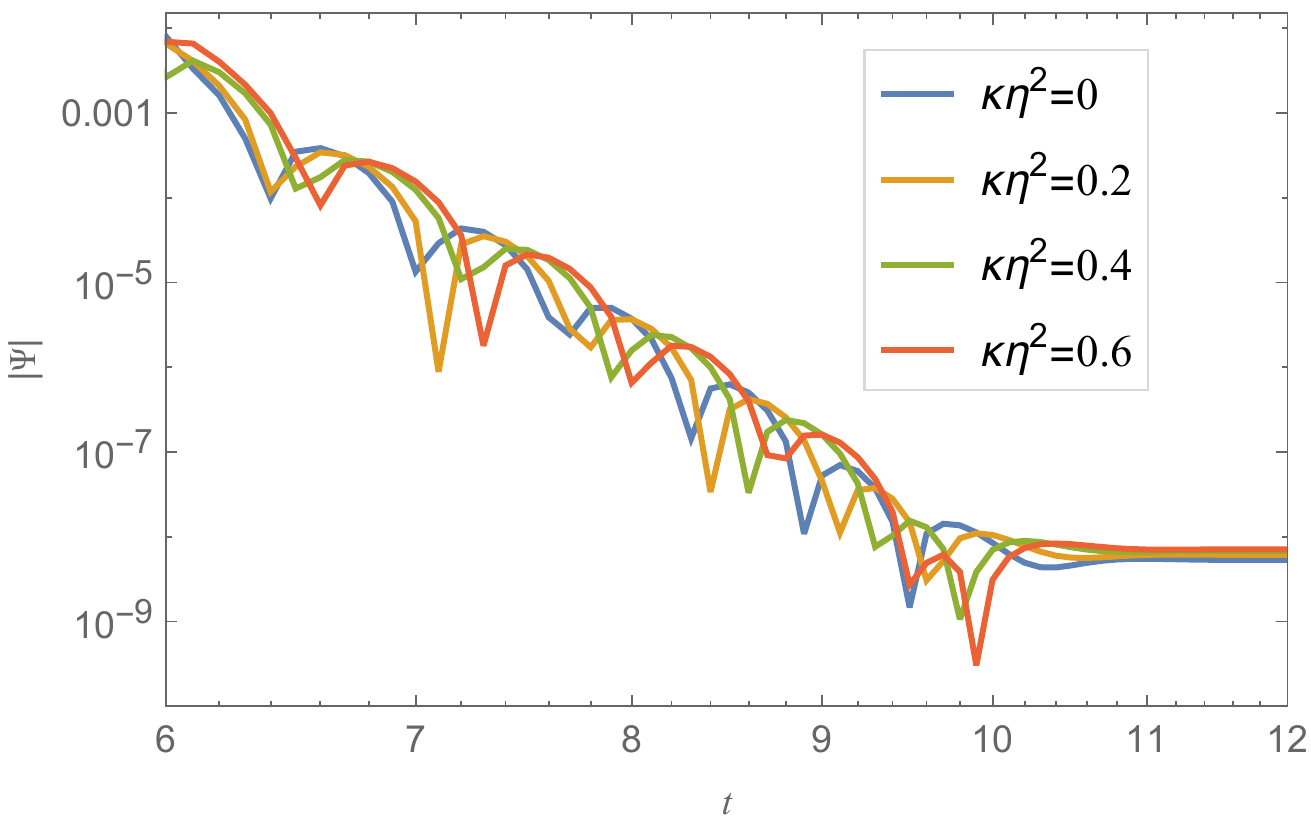}
    \caption{Temporal evolution of the scalar perturbations in the scenario with nonzero $\Lambda$,
        where $\kappa \eta ^{2}=0.2,l=2$ in the left panel
        and $L=0.5,l=2$ in the right panel.}
    \label{temp-ads}
\end{figure}

Fig. \ref{temp-ads} illustrates the temporal evolution of scalar perturbations
around the bumblebee black hole with a global monopole and nonzero $\Lambda$
for various values of $L$ and $\kappa\eta^2$.
The black hole is stable under massless scalar field perturbations,
as evident from the decay of the scalar field.
The left panel displays the evolution of the scalar perturbation for various $L$
with $\kappa\eta^2=0.2$ and $l = 2$.
The scalar field obviously decays slower for larger $L$,
agreeing with the results shown in Fig. \ref{Fig.QNM-ads}.
The right panel exhibits the evolution of scalar perturbation for various $\kappa\eta^2$
with $L = 0.5$ and $l = 2$.
The decay speed of the scalar field is very similar for different values of $\kappa\eta^2$.

\begin{figure}
    \centering
    \includegraphics[width=0.48\linewidth]{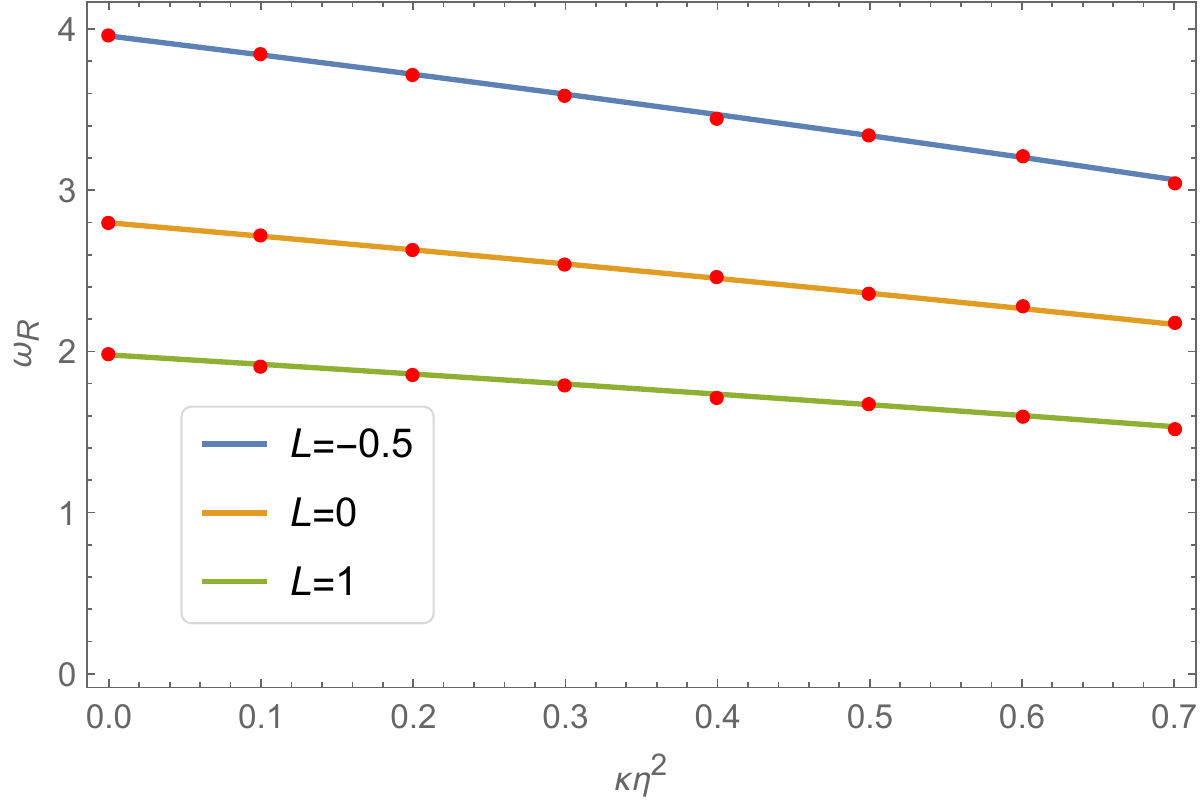}
    \includegraphics[width=0.48\linewidth]{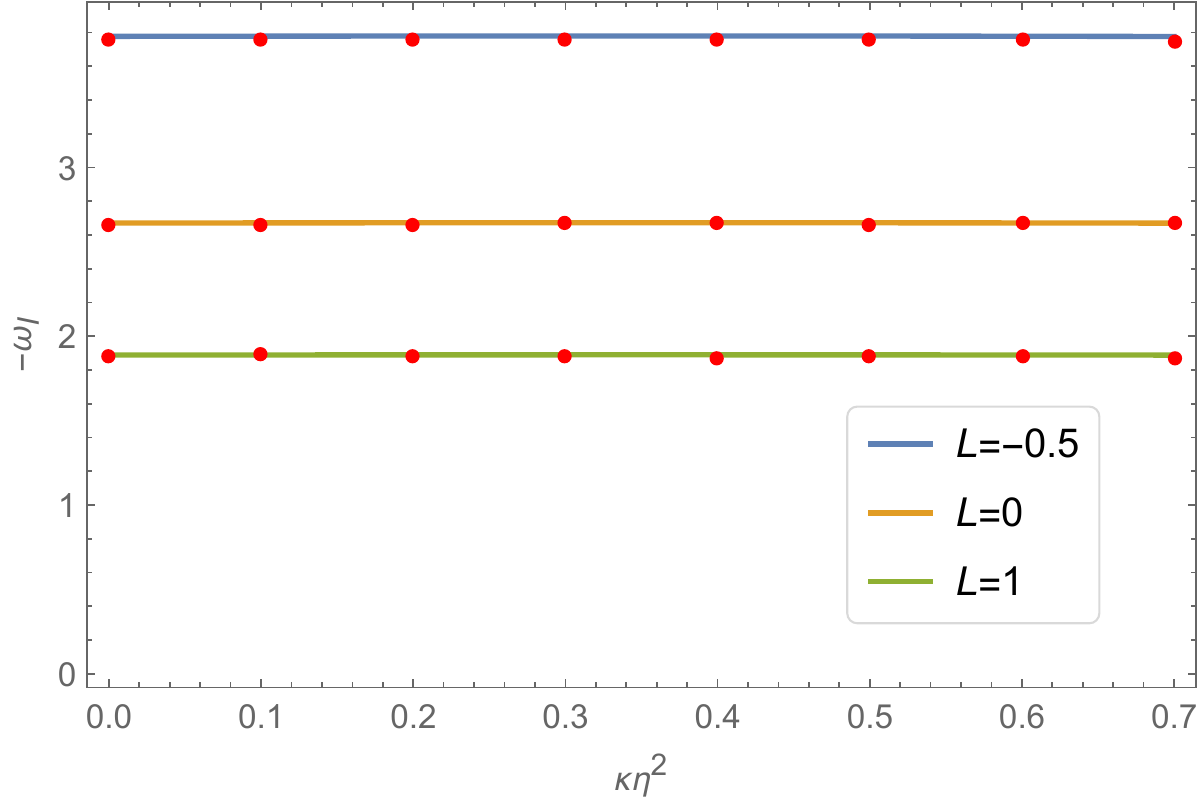}
    \caption{Comparison of the QNM frequencies obtained by the Horowitz–Hubeny method (solid lines)
        and those extracted from time domain analysis (red dots), with $\Lambda=-3,l=0$
        and $M$ chosen such that the horizon radius $r_+ = 1$.}
    \label{prony-ads}
\end{figure}

We again employ the Prony method to extract the QNM frequencies from the time domain profile.
To verify the robustness of our calculations,
in addition to the previously shown cases of $l=2$,
we use the cases of $l=0$ as the examples to cross check the results from frequency and time domains,
which are presented in Fig. \ref{prony-ads}.
The lines correspond to the Horowitz–Hubeny method,
and the red dots represent the frequencies extracted by the Prony method.
The results exhibit good agreement between the two methods,
confirming the reliability of our results.

\section{Conclusion and discussion}
\label{conclusion}
We have investigated the stability of the spherical bumblebee black holes with a global monopole in the two cases with a zero or a negative cosmological constant
by examining the QNMs of the black holes under the perturbations of a massless scalar field.
The analyses are performed in both frequency domain and time domain
by using different numerical methods.
We find that both families of black holes are stable under the perturbations of a massless scalar field.
In particular, the scalar field around the black holes decays slower for larger LSB parameter $L$.
On the other hand, the influence of the global monopole on the stability of two types of black holes is different.
For the black hole with zero $\Lambda$, the global monopole field prolongs the decay of the perturbation.
Whereas for the black hole with negative $\Lambda$, the effect of the global monopole on the stability is negligible.

Such a significant difference may be understood as follows.
QNM frequencies are eigenvalues $\omega$ that satisfy the boundary conditions
corresponding to the outgoing wave at spatial infinity and the ingoing wave at the horizon.
Thus, computing QNM frequencies is an eigenvalue problem that depends on the boundary condition at spatial infinity.
When $\Lambda=0$, the boundary condition \eqref{bc1} at spatial infinity
depends on both $L$ and $\kappa\eta^2$ through the tortoise coordinate \eqref{toordinate}.
Hence, both $L$ and $\kappa\eta^2$ affect QNM frequencies in this case.
However, when $\Lambda<0$, the boundary condition is a Dirichlet type
that is independent of $L$ and $\kappa\eta^2$ due to the divergent potential.
Therefore, $\kappa\eta^2$ has little effect on the QNM frequencies.
Nevertheless, $L$ still affects the QNM frequencies because it breaks the symmetry between space and time.
Moreover, the QNMs are derived by analyzing the outgoing wave signals of the perturbation,
which propagates differently depending on $L$
because the temporal-radial symmetry of the metric is broken.
The global monopole, on the other hand,
only manifests itself in a solid deficit angle in the spacetime,
which does not contribute to the breaking of the temporal-radial symmetry.
This can also be seen by rewritting the effective potential $U(r)$ in Eq. \eqref{potential}
for large $r$ as
\begin{equation}
    \begin{split}
        U(r) & =(1-\kappa \eta ^{2}-\frac{2M}{r}+r^{2} )\left [ \frac{l(l+1)}{r^2} +\frac{2M}{(1+L)r^3}+\frac{2}{1+L}   \right ]               \\
        & = \frac{2r^2}{1+L} +\left [ \frac{2(1-\kappa \eta ^{2})}{1+L} +l(l+1) \right ] -\frac{2M}{(1+L)r}+\mathcal{O} (\frac{1}{r^2} ).
    \end{split}
\end{equation}
One can see that $L$ contributes to the coefficient of the $r^2$ term,
while $\kappa\eta^2$ only affects the constant term.
Therefore, the global monopole does not alter the shape of $U(r)$ significantly at large $r$
and hence, the stability of the black hole is almost independent of the global monopole.

In the high energy regime, the wavelength of the massless scalar field is negligible compared to the horizon scale of the black hole.
Thus, the field follows the null geodesics, which are influenced by the global monopole and the LSB parameter\cite{Gullu:2020qzu}.
These factors may also affect the absorption and scattering cross sections of the bumblebee black holes.
This is an interesting topic for future research.
\section*{Acknowledgement}
\label{ackn}
This work is supported by the National Science Foundation of China under Grant No. 12105179.

%\appendix

\bibliography{ref}
\end{document}